\documentclass[aps,prl,twocolumn,amsmath,showpacs,letterpaper,floatfix]{revtex4}

\usepackage{graphicx}           
\usepackage{amsmath,amssymb}

\graphicspath{{images/}}         

\newcommand{\ket}[1]{\left| #1\right\rangle}

\newcommand{\iea}[0]{{\it et al.}}

\begin{document}

\title{Quantum memory for squeezed light}
\author{J\"urgen Appel\footnote{Present address: Niels Bohr Institute, Blegdamsvej 17 DK - 2100 K{\o}benhavn, Denmark.}}
\author{Eden Figueroa\footnote{The first two authors contributed equally to the work.}}
\author{Dmitry Korystov}
\author{M. Lobino}
\author{A. I. Lvovsky}

\affiliation{Institute for Quantum Information Science, University of
  Calgary, Calgary, Alberta T2N 1N4,
  Canada}
\homepage{http://www.iqis.org/}

\date{\today}

\begin{abstract}
We produce a 600-ns pulse of 1.86-dB squeezed vacuum at 795 nm in an optical parametric amplifier and store it in a rubidium vapor cell for 1 $\mu$s using electromagnetically induced transparency. The recovered pulse, analyzed using time-domain homodyne tomography, exhibits up to $0.21\pm 0.04$ dB of squeezing. We identify the factors leading to the degradation of squeezing and investigate the phase evolution of the atomic coherence during the storage interval.
\end{abstract}


\maketitle

\paragraph{Introduction} \label{sec:Introduction}

Memory for quantum states of light is a necessary component of quantum optical computers and is also required for the implementation of quantum repeaters
\cite{Briegel_Zoller} that would dramatically increase the range of quantum communication. There exists a variety of approaches to implementing quantum optical memory, for example off-resonant interaction of light with spin polarized atomic ensembles \cite{Julsgaard20041} and controlled reversible inhomogeneous broadening \cite{CRIB}. One of the most well-studied techniques is adiabatic transfer between optical quantum states and long-lived atomic superposition using electromagnetically
induced transparency (EIT) \cite{Fleisch_77}. This method, proposed in 2000 by Fleischhauer and Lukin \cite{Fleisch_2000}, has been experimentally demonstrated in 2001 with classical light pulses \cite{lukin01}. In 2005, storage and retrieval of single photons, prepared using the
Duan-Lukin-Cirac-Zoller protocol \cite{Duan20015}, was achieved
independently by Chaneli\`ere  \emph{et al.}
\cite{kuzmichnature} and Eisaman \emph{et al.} \cite{lukinnature}. These experiments have shown that a quantum state's nonclassical character is preserved after storage and retrieval.

In this letter, we demonstrate compatibility of the quantum memory technology with the continuous-variable domain of quantum optics by reporting, for the first time, storage and retrieval of the squeezed vacuum state in atomic rubidium vapor. We show that the light in the retrieved mode retains quadrature squeezing, albeit degraded by absorption and atomic decoherence. In addition, we demonstrate that the optical phase of the retrieved squeezed vacuum faithfully reproduces that of the input.

An important new feature of our experiment is that, rather than verifying a particular property of the retrieved ensemble, we perform full characterization of the input and retrieved states using pulsed, time-domain homodyne tomography \cite{smi93,UL,HTReview}. In this way, we obtain density matrices of both ensembles and can compare them, in particular, evaluate the memory fidelity. Our setup is thus universal in that it can be applied to testing quantum memory for any arbitrary quantum optical state.

Interaction of squeezed light with atoms under EIT conditions has been investigated by several research groups. Emergence of periodically-poled nonlinear materials enabled construction of narrowband parametric squeezed light sources resonant with atomic rubidium transitions \cite{Kozuma_OPO,PKLam_OPO,Quantech_OPO}. Propagation and slowdown of squeezed light in an EIT medium has been reported \cite{Kozuma_EIT_2004,Kozuma_cw_SL_EIT,Kozuma_slowdown} as well as storage of parametric fluorescence \cite{Kozuma_storage_PF}. Hsu \iea\  have investigated the effect of propagation through an EIT medium on the quadrature noise of the coherent state and found the atomic gas to generate a large excess noise at low sideband frequencies \cite{PKLam_noise}. This noise has been further studied theoretically and a set of benchmarks for quantum memory performance has been elaborated by H\'{e}tet \iea\  \cite{Hetet}.



\begin{figure}
 \includegraphics[width=0.9\columnwidth]{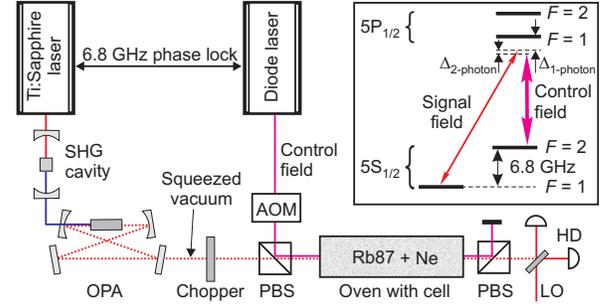}\\
 \caption{Experimental setup (SHG, second harmonic generation; PBS, polarizing
    beam splitter; LO, local oscillator; HD, homodyne detector;
    AOM, acousto optical modulator). The inset shows the atomic level configuration.}\label{setup}
\end{figure}


\paragraph{Experimental setup (Fig.~1)}
The master oscillator, a Coherent MBR-110 Ti:Saphire laser with a narrow spectral width
($\sim$40 kHz) and high long-term stability is tuned to the $^{87}$Rb $\ket{5S_{1/2},F=1}$--$\ket{5P_{1/2},F=1}$ transition at 795 nm.
Its second harmonic, produced by means of a TekhnoScan frequency doubling cavity, pumps our squeezed light source, a degenerate optical parametric amplifier (OPA), with a power of 40 mW. This device is similar to that described in Ref.~\cite{Quantech_OPO}, but employs a 20-mm periodically poled KTiOPO$_4$ crystal. When the signal output of the OPA is measured with a homodyne detector, it exhibits, on average, a 3.5 dB noise reduction with respect to the shot noise level (SNL) in the squeezed quadrature and 8 dB excess noise in the antisqueezed quadrature in a sideband range of at least 2~MHz.

Our goal is to observe storage of squeezed light, so the OPA output has to be chopped into microsecond pulses.
Because optical losses degrade squeezing, we avoid using electro- or acousto-optical modulators, and employ an ultra-fast mechanical chopper custom-made from a standard 3-inch computer hard disk. To its outer edge we attach another disk of a 9 mm diameter with a 50 $\mu$m slit (Thorlabs S50R) cut through its center. The hard disk is then positioned so that when rotating, the disk with the slit passes through the OPA output beam focused to a 25 $\mu$m size. 
For most of the 4-ms rotation period the squeezed vacuum passes
through the chopper wheel unobstructed. When the disk with the slit enters the beam it interrupts the light for about 60 $\mu$s, generates a 600 ns (FWHM) squeezed vacuum pulse and blocks the light for another 60 $\mu$s. An intensity waveform of a pulse produced by the chopper acting on a classical laser beam is given in Fig.~2(a).

The 5-mW EIT control field is provided by a separate diode laser. The frequencies of the signal and control fields are set to optimize, on one hand, the classical light storage efficiency, and on the other hand, transmission of squeezed light through the cell under EIT conditions. We found these requirements to be best fulfilled when the frequencies of both fields are red detuned from the center of the Doppler-broadened atomic line by $\Delta_\textrm{1-photon}=630$ MHz. In addition, the transmission of squeezing is improved when the fields are two-photon detuned from the 6834.68-MHz hyperfine splitting resonance by about $\Delta_\textrm{2-photon}=+0.54$ MHz. This is because the EIT resonance line is asymmetric, so such detuning permits better accommodation of the entire bandwidth of pulsed squeezed vacuum into the transparency window. The stability of $\Delta_\textrm{2-photon}$, which is crucial for the light storage procedure, is achieved by phase locking the diode laser to the master oscillator.

The rubidium vapor cell used for storage has 10 torr of neon buffer gas and is contained in a magnetically shielded oven heated to 65$^\circ$C. The control and signal fields are orthogonally linearly polarized which allows their combination and separation at the oven entrance and exit. The spatial modes of the control and signal fields are carefully matched to each other. The beam radius inside the cell is 600 $\mu$m.
Under these conditions, the lifetime of quantum optical memory is 1.3 $\mu$s.


The control field is on for most of the 4-ms experimental cycle. It is turned off, by means of an acousto-optical modulator, at time $t=1.02\ \mu$s (as defined by Fig.~2) when most of the signal pulse has entered the cell, and turned back on after the 1 $\mu$s storage period. After an additional 4 $\mu$s, when the stored signal has been fully retrieved, the control field is turned off again briefly At this time, neither squeezed vacuum nor control field are entering the homodyne detector so we can obtain a clean sample of the vacuum state quadrature noise. The intensity waveform associated with the storage and retrieval of a classical laser pulse is given by Fig.~2(b).

Upon separation from the control field \cite{footnote0}, the signal is subjected to homodyne detection using a local oscillator (LO) derived from the master oscillator. The LO phase is varied, with a period of 2.5 s, by means of a piezoelectric transducer. The homodyne detector uses two Hamamatsu S3883 photodiodes with 94$\%$ quantum efficiency and offers a bandwidth exceeding 6~MHz.
The signal from the homodyne detector, corresponding to the quantum quadrature noise of the detected field, is fed to a digital oscilloscope and a spectrum analyzer for simultaneous time- and frequency-domain measurements.

Figure 2(c) shows the pointwise variance of 100,000 oscilloscope traces registering the homodyne detector output during the storage/retrieval procedure. When the control field is off and the signal pulse has terminated (1.4 $\mu$s $<t<$ 2.0 $\mu$s), the signal is in the vacuum state, so the detector outputs shot noise. Prior to the beginning of the signal pulse ($t<0.4\ \mu$s), and upon completion of the retrieval ($t>$ 4 $\mu$s), when the control field is on, the shot noise is elevated by about 0.1 dB due to the Raman scattering of the control field \cite{PKLam_noise,footnote}. When the front of the squeezed vacuum pulse is transmitted through the cell (0.4 $\mu$s $<t<$ 1.4 $\mu$s), the phase-averaged quadrature noise is significantly higher than the SNL. When the control field is turned back on after the storage period (2 $\mu$s $<t<$ 3 $\mu$s), the noise level increases again due to the retrieval of the stored squeezed state.

\begin{figure}
 \includegraphics[width=0.7\columnwidth]{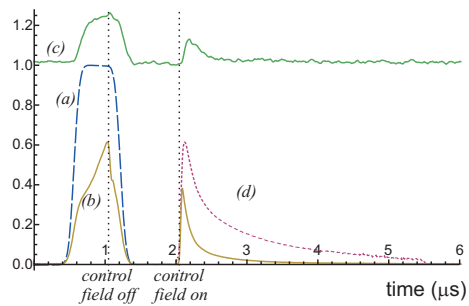}\\
 \caption{Time-dependent intensity of the classical input pulse without storage (a), and with storage (b). The $y$-axis units are arbitrary, but the intensity ratio between the fields is preserved. Line (c) shows time-domain pointwise variances of the phase-averaged homodyne detector photocurrent, in the units of SNL. Curve (d) is the temporal mode $f(t)$ in which the retrieved state is reconstructed.}
\end{figure}

\paragraph{Quantum state reconstruction}

We characterize the states of the electromagnetic field modes corresponding to the input and retrieved pulses. We define the temporal amplitude shape $f(t)$ of these modes by the square root of the associated classical intensity waveforms [Fig.~2(d)]. We use a continuous LO, and post-process the homodyne photocurrent by multiplying it by $f(t)$ and subsequently integrating over time \cite{tempfiltering}. In this fashion, we process the photocurrent from 100,000 retrieved pulses and obtain a set of values that are proportional to the field quadrature noise samples of the retrieved state. The proportionality coefficient is determined from the variance of 100,000 quadrature measurements of the vacuum state acquired in the same temporal mode.

Quantum state reconstruction requires knowing the LO phase values associated with each quadrature sample. To this end, we use a spectrum analyzer to continuously acquire the homodyne detector current in the zero-span mode at a 500 kHz sideband with a resolution bandwidth of 30 kHz and a sweep time of 2.5 seconds. Because this acquisition is relatively slow, and because the chopper
blocks the signal for less than 2\% of its rotation period, the chopper has very little effect on the acquired signal, which corresponds to the phase-dependent quadrature noise of the squeezed vacuum generated by the OPA. The LO phase is determined from this signal.

We apply the iterative likelihood-maximization algorithm \cite{Lvovsky_MLR} to reconstruct the input and retrieved states from  100,000 quadrature-phase pairs obtained in this manner [Fig.~3(a)]. This procedure yields the density matrices in the Fock basis [Fig.~3(b)] from which the Wigner function of the state in question is calculated [Fig.~3(c)] as well as the phase-dependent behavior of the quadrature noise [Fig.~3(d)].

\begin{figure}[t]
    \includegraphics[width=0.9\columnwidth]{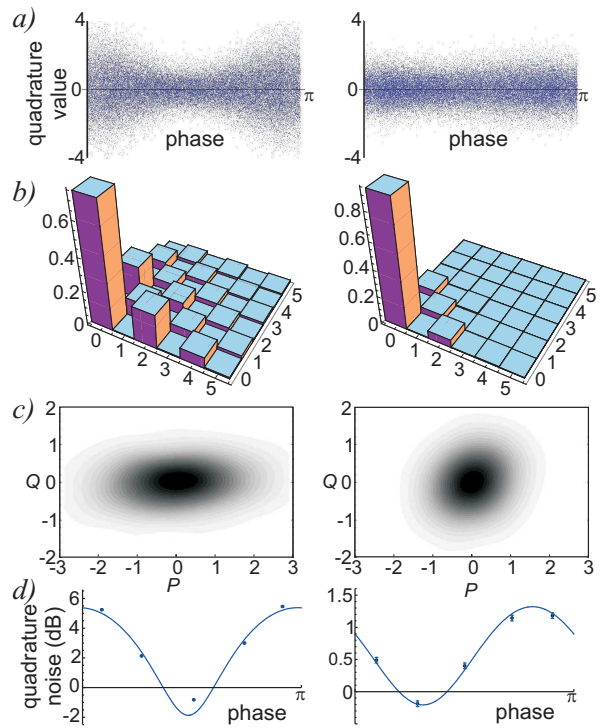}
  \caption{Quantum state of the input (left column) and retrieved (right column) states. Raw samples of phase-dependent quadrature noise (a), maximum-likelihood reconstruction of the density matrices in the Fock basis (absolute values, b), Wigner functions (c) and quadrature noise variances (d) are displayed. In (d), the solid lines are calculated from the density matrices while the points with error bars show variances of binned quadrature values, as discussed in the text.}
\end{figure}

Both the input and retrieved states are, to a large degree of accuracy, squeezed thermal states, and both exhibit the minimum quadrature noise at the level below the standard quantum limit. The squeezing for the input state is 1.86~dB (corresponding to 0.65~SNL on a linear scale) and for the retrieved state it is 0.21~dB (0.95~SNL). The corresponding antisqueezing levels amount to 5.38~dB (3.45~SNL) and 1.32~dB (1.36~SNL), respectively. The uncertainty of these values is estimated as described in Ref.~\cite{Lvovsky_MLR} as 0.03~dB. We see that the time-domain measurement of the input state already shows significant degradation of squeezing as compared to the frequency domain. This happens because (1) time-domain reconstruction includes low-frequency sidebands, where the squeezing is reduced and additional technical noise is present and (2) in the shoulders of the pulse, the slit edges clip the beam. The squeezing in the retrieved state is further reduced compared to the input. Again, two main factors are at play here: the storage efficiency (energy ratio of the input and retrieved pulses), which can be estimated from the classical data to be about 15\%, and the Raman excess noise.

We also verify the presence of squeezing in the retrieved mode by direct calculation of the phase-dependent quadrature variance. To this end, we partition the acquired quadrature set into 5 bins of $N=20,000$ points according to phases and evaluate the mean square variance within each bin [Fig.~3(d)]. One of the bins exhibits variance of $(0.187\pm0.043)$ dB below the SNL. The margin of error is evaluated here as the statistical error of estimating the width $\sigma$ of a Gaussian distribution from $N$ samples, and equals $\sigma \sqrt{2/N}$ \cite{Tracking}.

We perform two additional tests of our interpretation of this experiment as a demonstration of quantum memory for squeezed light. First, we store vacuum and analyze the quantum state in the retrieved optical mode. As expected, we detect a state with almost phase-independent quadrature noise, elevated by 0.1 dB with respect to the vacuum due to Raman scattering of the control field. Second, we store the squeezed state, but do not turn the control field back on for retrieval. We acquire and analyze the optical state in the mode defined by $f(t)$ and find it to be almost exactly vacuum, as expected.

\paragraph{Performance evaluation}
We quantify the performance of our memory setup in terms of fidelity, which, for the input and output quantum states $\hat\rho_{in}$ and $\hat\rho_{retr}$ is defined as $F={\rm Tr}[(\hat\rho^{1/2}_{in}\hat\rho_{retr}\hat\rho^{1/2}_{in})^{1/2}]^2$. Comparing the input and retrieved states, we determine $F=0.89$. This value is significantly higher than the ``classical'' fidelity of 0.74, defined by adding two units of the vacuum noise to the input state \cite{KimbleFidelity}. The same fidelity of 0.74 would be obtained in a ``memory'' procedure where the input state is replaced by the vacuum.

The reader should be cautioned that the obtained fidelity value is high mainly due to a large vacuum component in both $\hat\rho_{in}$ and $\hat\rho_{retr}$. An alternative figure of merit for our interface can be obtained by evaluating the degrees of nonclassicality of both states in terms of their entanglement potentials \cite{Asboth}, equalling 0.309 for $\hat\rho_{in}$ and 0.036 for $\hat\rho_{retr}$. We see that only a small fraction of the input state's nonclassicality is transferred to the retrieved pulse.

\paragraph{Phase evolution}
We use our setup to study the evolution of the
atomic coherence during the storage time with respect to that of the laser fields. If the signal and control field are not in exact two-photon resonance with respect to the atomic hyperfine splitting, the atomic and optical systems evolve with different frequencies. As a result, the retrieved state's phase, measured with respect to the local oscillator, is different from the input by
\begin{equation}
\phi = 2 \pi\Delta_\textrm{2-photon}\, \tau_{\rm storage}, \label{phasedep}
\end{equation}
 so we expect a linear behavior if either the two-photon detuning or the storage time are changed. We verify this dependence, reconstructing the retrieved state at different values of these parameters. This reconstruction yields phase values modulo $\pi$, so in order to fit the theoretical prediction (\ref{phasedep}) we add multiples of $\pi$ to each point. The result of this procedure is shown in Fig.~4 and exhibits good agreement with the theory.

\begin{figure}[t]
    \includegraphics[width=0.9\columnwidth]{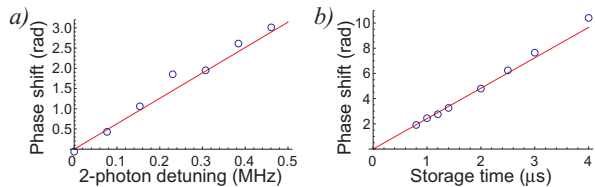}
  \caption{Phase dependence of the reconstructed squeezed state with respect to the local oscillator as a function of the two-photon detuning with a constant $\tau_{\rm storage}=1\ \mu$s (a) and as a function of the storage time with a constant $\Delta_\textrm{2-photon}=384$ kHz (b). The solid lines correspond to Eq.~(\ref{phasedep})}.
\end{figure}

\paragraph{Summary}
We have demonstrated
storage and retrieval of pulsed squeezed light using EIT. 
For the first time, complete tomographic characterization of the quantum state after the storage procedure has been performed, making our setup a universal testbed for a generic quantum optical memory system.

\paragraph{Acknowledgements}
We gratefully acknowledge M. Eisaman, P. Kolchin, F. Vewinger, K.-P.
Marzlin and B. C. Sanders for fruitful discussions, as well as G. G\"unter and Y. Takeno for their help in the laboratory. This work was
supported by NSERC, CIAR, iCORE, AIF, CFI and Quantum\emph{Works}.

\paragraph{Note added} While this paper was being prepared for submission, we have become aware of the work \cite{Kozuma_mem}, where storage and retrieval of squeezed vacuum in a cold atom ensemble has been demonstrated, albeit without tomographic reconstruction.



\begin{thebibliography}{13}
\bibitem{footnote0} The extinction ratio for separating the control and signal fields is about $10^{-4}$. The control field and the local oscillator are of different frequencies, so a small fraction of the control field transmitted through the polarizers has no effect on the state reconstruction.
\bibitem{footnote} This noise level is greatly reduced compared to that observed in Ref.~\cite{PKLam_noise}. We believe this to be mainly due to the presence of a buffer gas in the cell.
\end{thebibliography}
\end{document}